\documentclass{amsart}

\usepackage{a4wide, enumerate, amsmath, amsfonts, amssymb, amsthm, wasysym, graphics, graphicx, xcolor, url, hyperref, hypcap, multirow, algorithm, algorithmic}
\hypersetup{colorlinks=true, citecolor=darkblue, linkcolor=darkblue}
\graphicspath{{figures/}}

\title{On topological and geometric $(19_4)$\,configurations}

\author[J.~Bokowski]{J\"urgen Bokowski}
\address[J.~Bokowski]{Technische Universit\"at Darmstadt}
\email{juergen.bokowski@gmail.com}

\author[V.~Pilaud]{Vincent Pilaud$^{\ddagger}$} 
\address[V.~Pilaud]{CNRS \& LIX, \'Ecole Polytechnique, Palaiseau}
\email{vincent.pilaud@lix.polytechnique.fr}
\urladdr{http://www.lix.polytechnique.fr/~pilaud/}
\thanks{$^\ddagger$VP was supported by the spanish MICINN grant MTM2011-22792, by the French ANR grant EGOS 12 JS02 002 01, and by the European Research Project ExploreMaps~(ERC~StG~208471).}

\newtheorem{theorem}{Theorem}[section]

\newtheorem{fact}[]{Result}

\theoremstyle{definition}
\newtheorem{example}[]{Example}
\newtheorem{remark}[theorem]{Remark}

\newcommand{\R}{\mathbb{R}}
\newcommand{\Z}{\mathbb{Z}}

\newcommand{\bP}{\mathbb{P}}
\newcommand{\bE}{\mathbb{E}}
\newcommand{\bI}{\mathbb{I}}

\newcommand{\ssm}{\smallsetminus}
\newcommand{\eqdef}{\mbox{\,\raisebox{0.2ex}{\scriptsize\ensuremath{\mathrm:}}\ensuremath{=}\,}} % :=

\newcommand{\ie}{\textit{i.e.}~} % id est
\newcommand{\eg}{\textit{e.g.}~} % exempli gratia
\newcommand{\etc}{etc.} % etc
\newcommand{\pl}{point\,--\,line} % point-line relation

\newcommand{\conf}[2]{\mbox{$(#1_#2)$\,confi}\-gu\-ra\-tion}

\newcommand{\Maple}{\textsc{maple}}
\newcommand{\Java}{\textsc{java}}

\newcommand{\Cinderella}{\textsc{cinderella}}

\definecolor{darkblue}{rgb}{0,0,0.7}
\newcommand{\darkblue}{\color{darkblue}}
\newcommand{\defn}[1]{\emph{\darkblue #1}}
\graphicspath{{figures/}}
\renewcommand{\paragraph}[1]{\vspace{.3cm}\noindent{\bf #1} ---}
\newcommand{\ssep}{\vspace*{.3cm}\hspace*{0pt plus 1fill}\rule{5 cm}{.5 pt}\vspace{.3cm}\hspace*{0pt plus 1fill}}

\usepackage{todonotes}

\begin{document}

\maketitle

\begin{abstract}
An \conf{n}{k} is a set of~$n$ points and~$n$ lines such that each point lies on~$k$ lines while each line contains~$k$ points. The configuration is geometric, topological, or combinatorial depending on whether lines are considered to be straight lines, pseudolines, or just combinatorial lines. The existence and enumeration of \conf{n}{k}s for a given~$k$ has been subject to active research. A current front of research concerns geometric \conf{n}{4}s: it is now known that geometric \conf{n}{4}s exist for all~$n \ge 18$, apart from sporadic exceptional cases. In this paper, we settle by computational techniques the first open case of \conf{19}{4}s: we obtain all topological \conf{19}{4}s among which none are geometrically realizable.
\end{abstract}

%%%%%%%%%%%%%%%%%%%%%%%%%%%%%%
%%%%%%%%%%%%%%%%%%%%%%%%%%%%%%

\section{Introduction}

\begin{figure}[h]
  \centerline{\includegraphics[width=.8\textwidth]{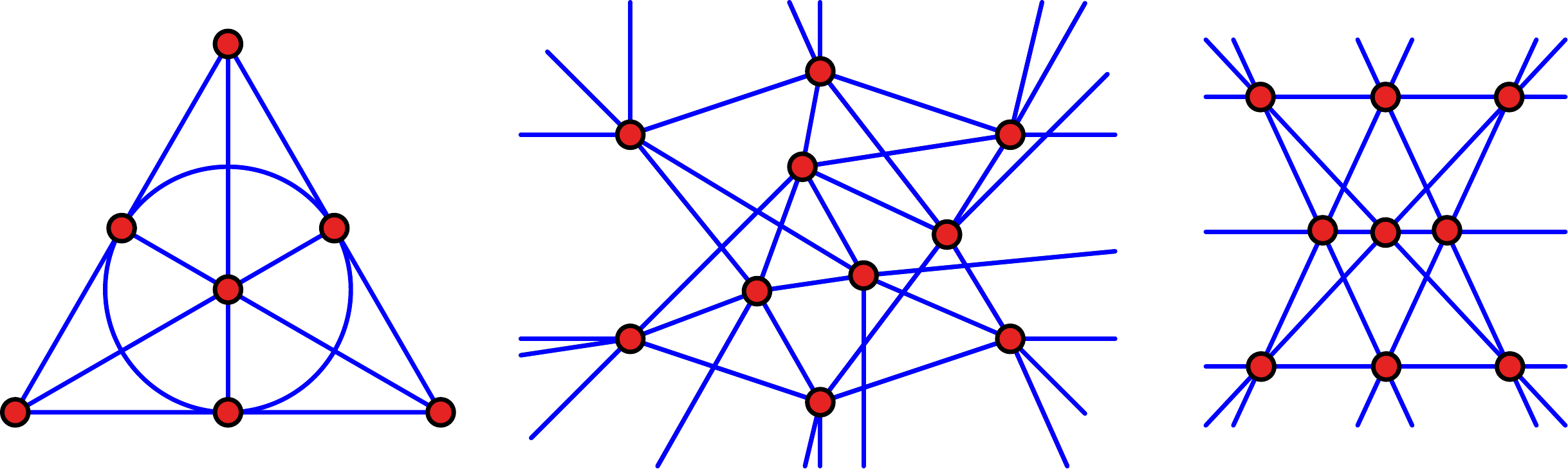}}
  \caption{Fano's combinatorial \conf{7}{3} (left), Kantor's topological \conf{10}{3} (center), and Pappus' geometric \conf{9}{3} (right).}
  \label{fig:famousConfigurations}
\end{figure}

An \defn{\conf{n}{k}} is formed by a set~$P$ of~$n$ \defn{points} and a set~$L$ of~$n$ \defn{lines} such that each point of~$P$ lies on precisely~$k$ lines of~$L$ while each line of~$L$ contains precisely~$k$ points of~$P$. The different possible meanings for points and lines define different notions of configurations:
\begin{enumerate}[(i)]
\item For \defn{geometric configurations}, points and lines are ordinary points and lines in the real projective plane~$\bP$.
\item For \defn{topological configurations}, points are ordinary points in~$\bP$, but lines are \defn{pseudolines}, \ie non-separating simple closed curves of~$\bP$ which cross pairwise precisely once.
\item For \defn{combinatorial configurations}, we just consider abstract points and lines, together with an incidence relation such that no two distinct points are incident to two distinct lines. Equivalently, combinatorial \conf{n}{k}s can be described as $k$-regular bipartite graphs on $2n$ vertices with girth at least~$6$.
\end{enumerate}
Famous examples of such configurations are represented in Figure~\ref{fig:famousConfigurations}. These examples reflect the long history of \conf{n}{k}s, and their connections to projective incidence theorems and realizability problems. A detailed survey on configurations including historical perspectives and careful references to the literature can be found in the recent monograph of B.~Gr\"unbaum~\cite{Grunbaum1}.

\begin{figure}
  \centerline{\includegraphics[height=6.8cm]{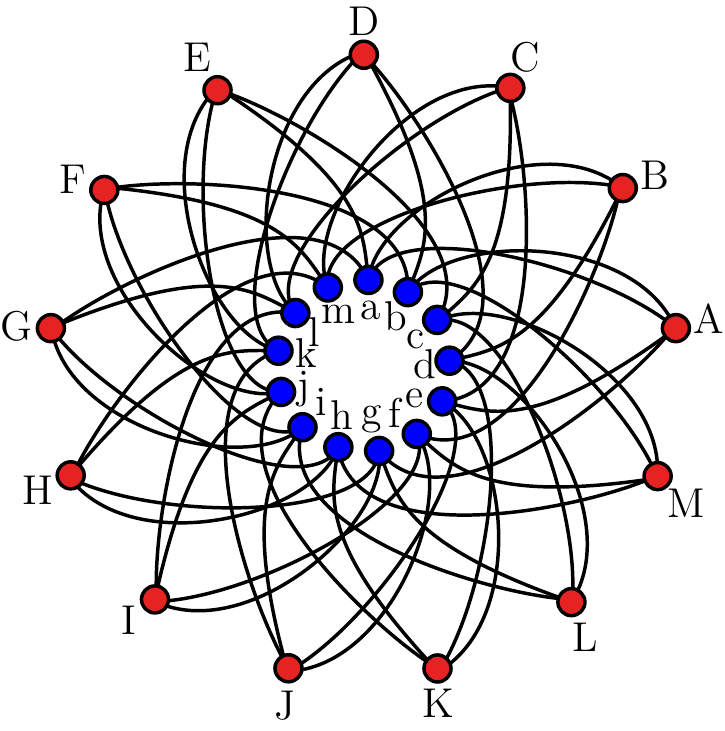} \; \includegraphics[height=6.8cm]{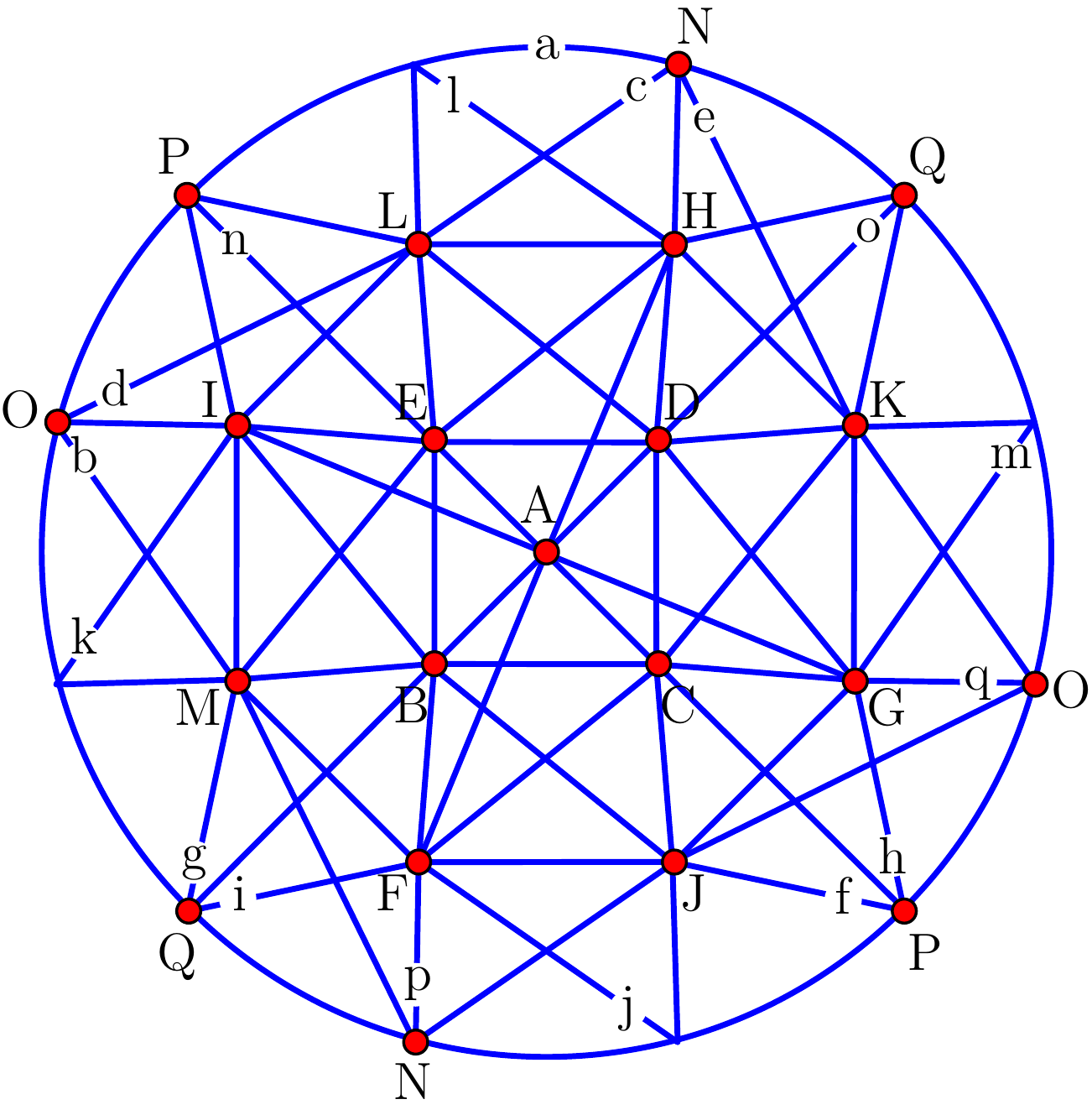}}
  \caption{The incidence graph (Levi graph) of the first combinatorial \conf{13}{4} (left) and the first topological \conf{17}{4}~\cite{BokowskiSchewe2}~(right).}
  \label{fig:smallConfs}
\end{figure}

\begin{figure}[h]
  \centerline{\includegraphics[height=6.2cm]{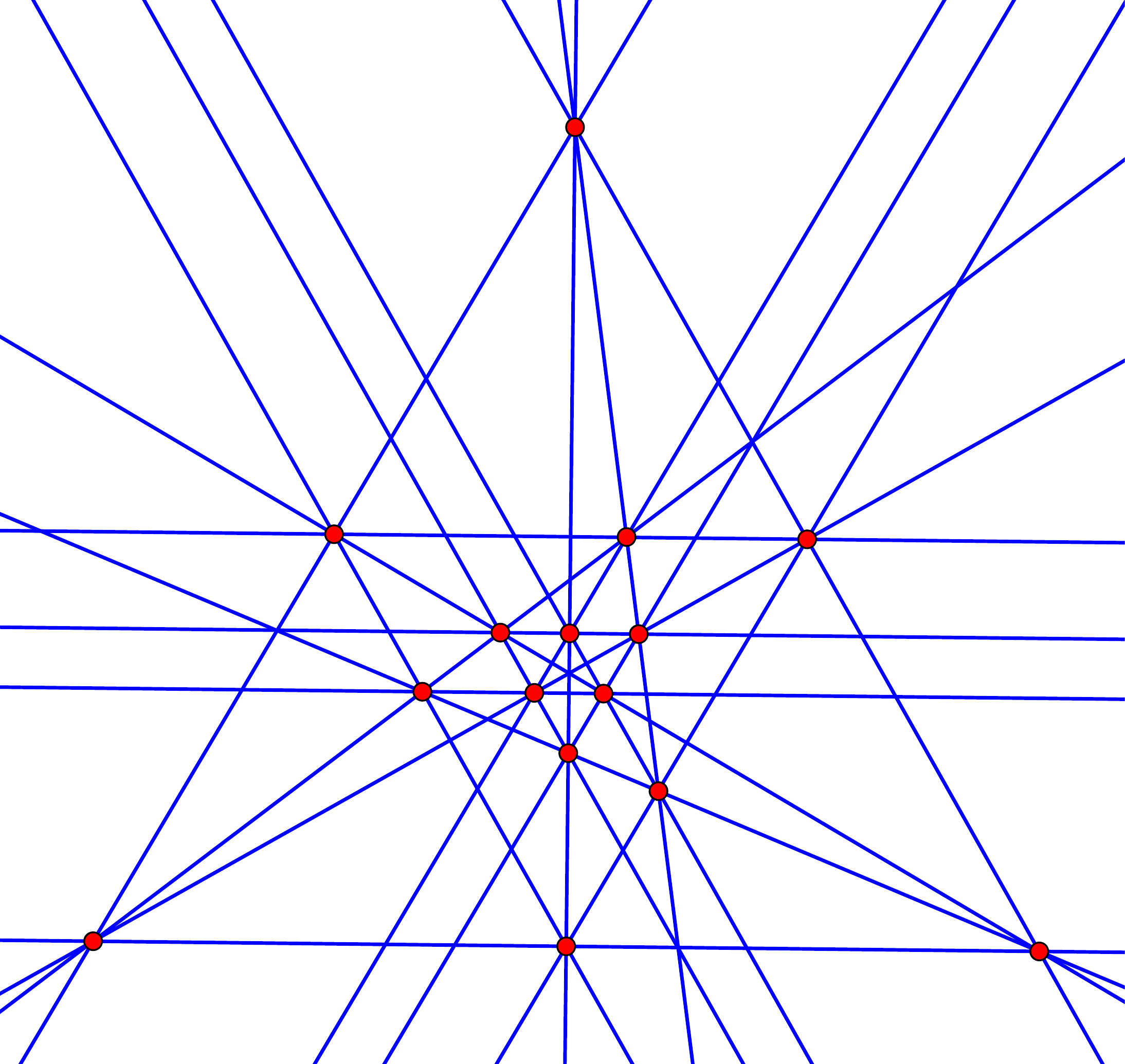} \qquad \includegraphics[height=6.2cm]{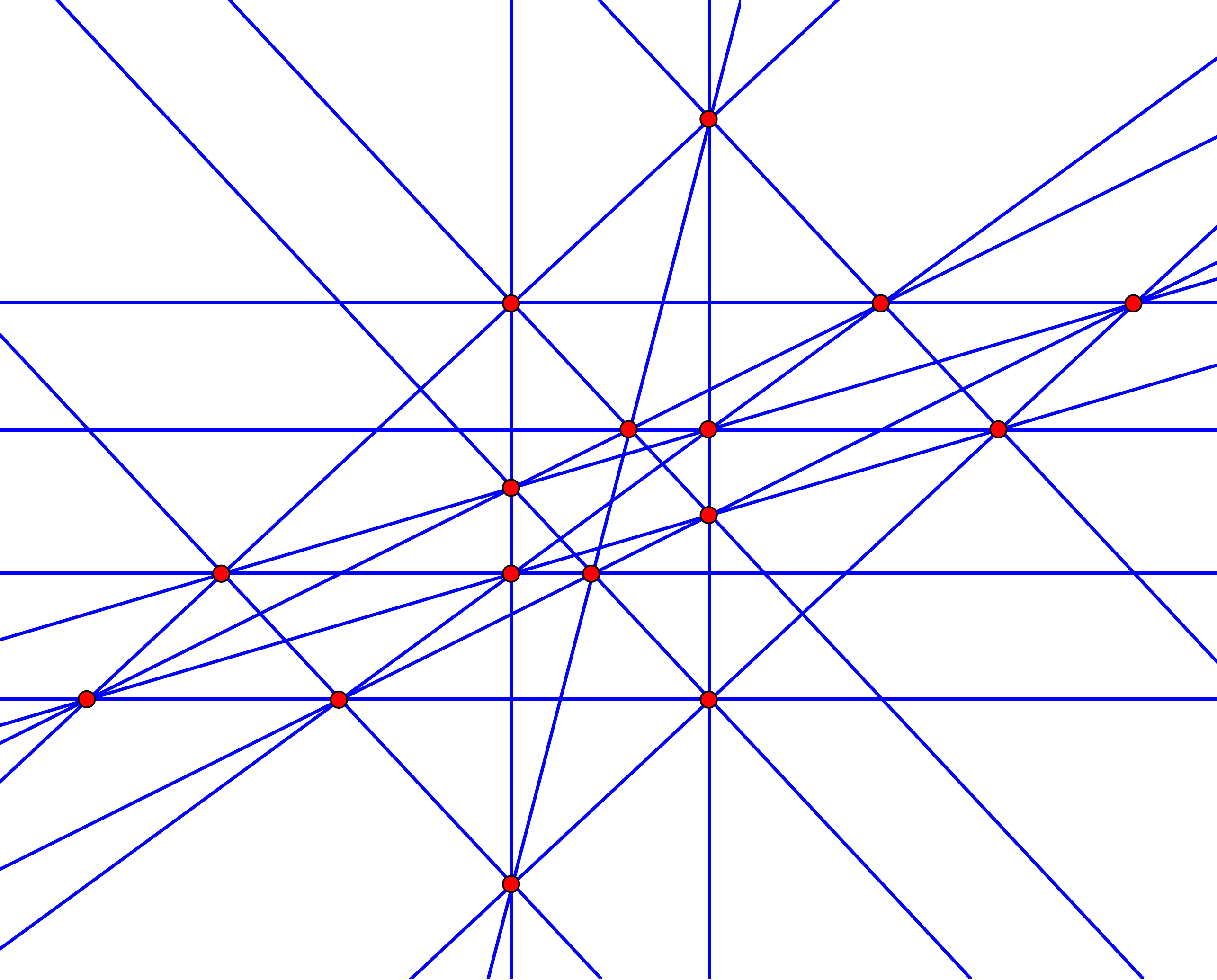}}
  \caption{The two geometric \conf{18}{4}s~\cite{BokowskiSchewe1, BokowskiPilaud-topologicalConfigurations}.}
  \label{fig:18_4}
\end{figure}

The first question on configurations raised in B.~Gr\"unbaum's monograph is to describe, for a fixed integer~$k$, for which values of~$n$ do combinatorial, topological, and geometric \conf{n}{k}s exist. This question is completely settled for~$k \le 3$, extensively studied for~$k = 4$, and still widely open for~$k \ge 5$. When~$k = 4$, the existence of geometric \conf{n}{4}s has been proved for all sufficiently large integers~$n$ by various geometric constructions, most of them using non-trivial symmetry groups. These constructions are surveyed in~\cite[Chapter~3]{Grunbaum1}. The remaining integers~$n$ have been treated individually, with ad-hoc constructions or arguments to prove or disprove the existence of \conf{n}{4}s. The current state of knowledge is the following: combinatorial \conf{n}{4}s exist iff~$n \ge 13$, topological \conf{n}{4}s exist iff~$n \ge 17$ \cite{BokowskiSchewe1, BokowskiGrunbaumSchewe} and geometric \conf{n}{4}s exist iff $n\ge18$~\cite{Grunbaum2, Grunbaum3, Grunbaum4, BokowskiSchewe2}, with the possible exceptions\footnote{In fact, there are currently only $3$ remaining cases. Indeed, case~$n = 19$ is treated in detail in this paper, while cases~$n = 37$ and $43$, as well as several by-products of our investigation on small \conf{n}{k}s, will be discussed in a separate paper, to keep the present paper short and focused on topological and geometric \conf{19}{4}s.} of $n = 19$, $22$, $23$, $26$, $37$ and~$43$. To illustrate these results, we have represented in Figures~\ref{fig:smallConfs} and~\ref{fig:18_4} the first examples of \conf{n}{4}s: Figure~\ref{fig:smallConfs} shows the incidence relation of the first combinatorial \conf{13}{4} and the first topological \conf{17}{4} (antipodal points of the circle are identified), while Figure~\ref{fig:18_4} shows the only two geometric \conf{18}{4}s~\cite{BokowskiSchewe1, BokowskiPilaud-topologicalConfigurations} (some points are at infinity to obtain more symmetric pictures).

This paper settles the case of \conf{19}{4}s. All $269\,224\,653$ combinatorial \conf{19}{4}s were recently enumerated in~\cite{PaezOsunaSanAugustinChi}. It should however be clear that searching for topologically or geometrically realizable configurations in this list would be like looking for a needle in a haystack. We have instead developed and implemented in~\cite{BokowskiPilaud-topologicalConfigurations} an algorithm to generate directly all topological \conf{19}{4}s up to combinatorial equivalence, which does not start from the list of all combinatorial \conf{19}{4}s. Using this algorithm as a black box, we report in Section~\ref{sec:topological} on the list of all~$4028$ topological \conf{19}{4}s, with a particular attention to their isomorphism group. We then present in Section~\ref{sec:geometric} the so-called construction sequence method which we use on the one hand to search for subconfigurations in a configuration and on the other hand to test the geometric realizability of a configuration. Using this method, we surprisingly conclude that there is no geometric \conf{19}{4}.

%%%%%%%%%%%%%%%%%%%%%%%%%%%%%%
%%%%%%%%%%%%%%%%%%%%%%%%%%%%%%

\section{Topological \conf{19}{4}s}
\label{sec:topological}

We developed in~\cite{BokowskiPilaud-topologicalConfigurations} an algorithm to enumerate directly all topological \conf{n}{k}s without enumerating first the combinatorial \conf{n}{k}s. This algorithm, implemented in \Java, enumerates all topological \conf{19}{4}s in approximately two weeks\footnote{Computation time on a 2.4 GHz Intel Core 2 Duo processor with 4Go of RAM.}:

\begin{fact}[\cite{BokowskiPilaud-topologicalConfigurations}]
There are precisely $4\,028$ topological \conf{19}{4}s up to combinatorial equivalence. Among them, $222$ are self-dual.
\end{fact}

Studying this list, we can already answer B.~Gr\"unbaum's problem to find symmetric topological \conf{19}{4}s~\cite[p.\,169, Question~5]{Grunbaum1}. First, the combinatorial automorphism groups of these configurations are distributed as follows:
\medskip
\begin{center}
\renewcommand{\arraystretch}{1.3}
\begin{tabular}{c||c|c|c|c|c||c}
  group $G$ & $1$ & $\Z_2$ & $\Z_2 \times \Z_2$ & $\Z_2 \times \Z_2 \times \Z_2$ & $D_8$ & Total \\
  \hline
  number of \conf{19}{4}s & \multirow{2}{*}{$3\,726$} & \multirow{2}{*}{$283$} & \multirow{2}{*}{$14$} & \multirow{2}{*}{$2$} & \multirow{2}{*}{$3$} & \multirow{2}{*}{$4028$} \\[-.2cm]
  with automorphism group~$G$ & & & & &
\end{tabular}
\end{center}
\medskip
By \defn{combinatorial automorphisms} of a configuration~$(P,L)$ we mean the automorphisms of its Levi graph. Remember that the \defn{Levi graph} of a \pl{} configuration~$(P,L)$ is the bipartite graph whose nodes are the elements of $P \sqcup L$ and whose edges relate incident elements. An example is given in Figure~\ref{fig:smallConfs}\,(left). An automorphism of the Levi graph of a configuration~$(P,L)$ is a \defn{preserving automorphism} of the configuration when it fixes the two maximal independent sets $P$ and $L$, and a \defn{self-duality} of the configuration if it exchanges $P$ and $L$. For example, in Figure~\ref{fig:smallConfs}\,(left), the rotation of angle $2\pi/13$ is a preserving automorphism, while the transformation~(A,a)(B,b)\,\dots\,(L,l)(M,m) is a self-duality.

We then tried to realize the five configurations whose combinatorial automorphism group has order~$8$ in such a way that their preserving automorphism are realized as isometries, and that the self-dualities are self-polarities of the configuration. A \defn{self-polarity} of a topological configuration $(P,L)$ is a self-duality $^* : (P,L) \mapsto (L^*, P^*)$ which respects cyclic orders: if the points ${p_1, \dots, p_k \in P}$ appear in cyclic order along a pseudoline $\ell \in L$, then the dual pseudolines $p_1^*,\dots,p_k^*\in P^*$ must appear in cyclic order around the dual point $\ell^* \in L^*$, and similarly if the lines $\ell_1, \dots, \ell_k \in L$ appear in cyclic order around a point $p\in P$, then the dual points $\ell_1^*, \dots, \ell_k^* \in L^*$ must appear in cyclic order along the dual line~$p^* \in P^*$. For example, the transformation~(A,a)(B,b)\,\dots\,(R,r)(S,s) is a self-polarity of the topological configuration of Figure~\ref{fig:smallConfs}\,(right).

\begin{figure}
  \centerline{\includegraphics[scale=.7]{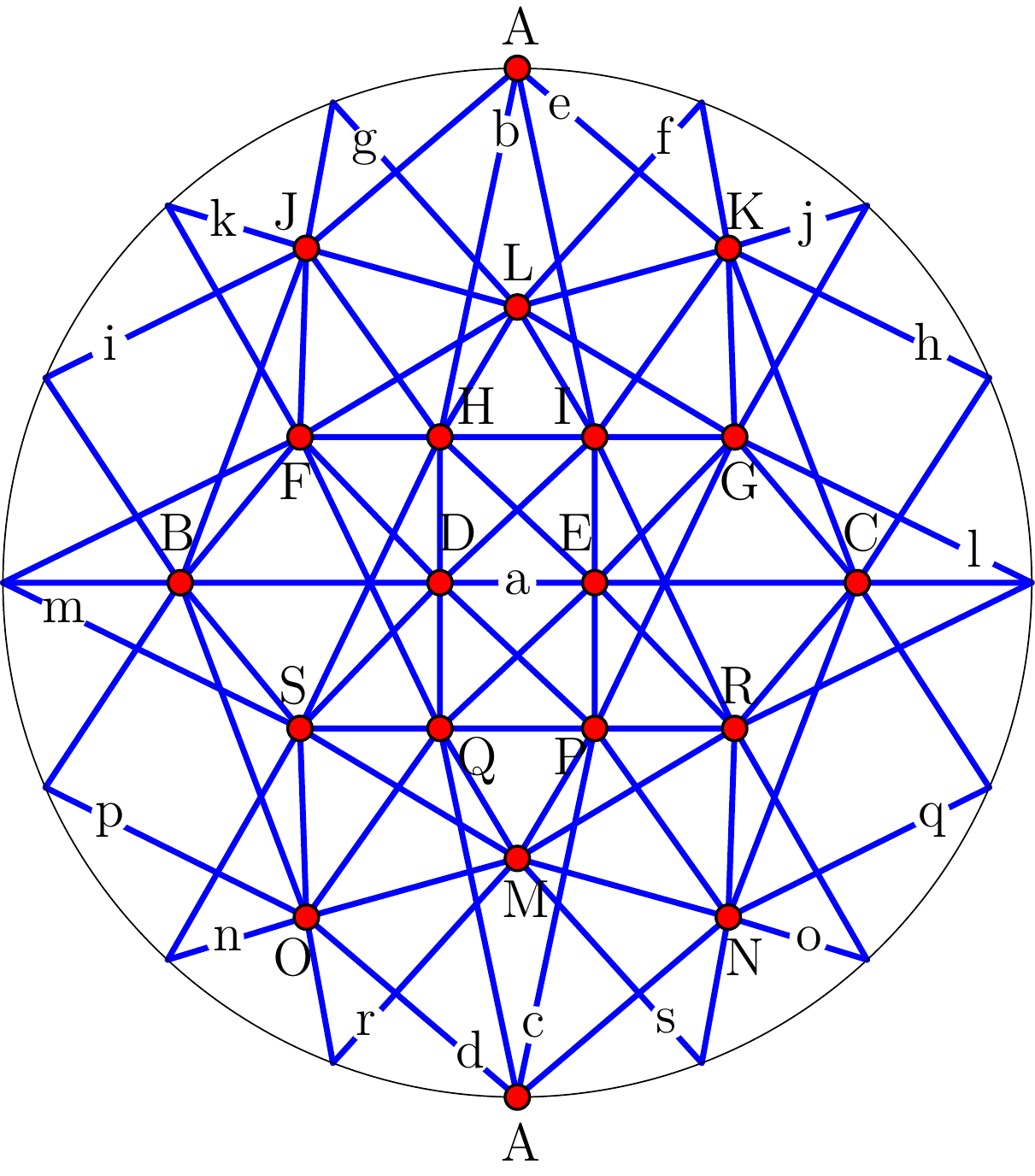}}
  \caption{A topological \conf{19}{4} whose isometry group is $\Z_2 \times \Z_2$, and with an additional self-polarity.}
  \label{fig:configuration_19_4_1}
\end{figure}

The most symmetric topological \conf{19}{4} that we obtained is represented in Figure~\ref{fig:configuration_19_4_1}. Its \pl{} incidences  are given by the following table:
\medskip
\begin{center}
\begin{tabular}{c|cccccccccccccccccccc}
  lines 		& a & b & c & d & e & f & g & h & i & j & k & l & m & n & o & p & q & r & s \\
  \hline
  			& B & A & A & A & A & H & I & B & C & D & E & F & P & E & D & C & B & P & Q \\
  points 	& C & H & I & J & K & L & L & F & G & I & H & G & Q & Q & P & R & S & M & M \\
  in lines 	& E & D & E & B & C & O & N & L & L & K & J & I & S & O & N & M & M & K & J\\
    			& D & Q & P & O & N & S & R & K & J & S & R & H & R & G & F & O & N & G & F
\end{tabular}
\end{center}
% [[1,2,4,3],[0,7,3,16],[0,8,4,15],[0,9,1,14],[0,10,2,13],[7,11,14,18],[8,11,13,17],[1,5,11,10],[2,6,11,9],[3,8,10,18],[4,7,9,17],[5,6,8,7],[15,16,18,17],[4,16,14,6],[3,15,13,5],[2,17,12,14],[1,18,12,13],[15,12,10,6],[16,12,9,5]]
%
\bigskip
We have labeled the points and pseudolines of this configuration in such a way that:
\begin{itemize}
\item the action on the points of the vertical and horizontal reflexions of the picture are respectively given by the permutations
\begin{center}
(A)(B,C)(D,E)(F,G)(H,I)(J,K)(L)(M)(N,O)(P,Q)(R,S) and \\
(A)(B)(C)(D)(E)(F,S)(G,R)(H,Q)(I,P)(J,O)(K,N)(L,M);
\end{center}
\item the permutation (A,a)(B,b)\,\dots\,(R,r)(S,s) is a self-polarity.
\end{itemize}
The combinatorial automorphism group of the configuration is isomorphic to $\Z_2 \times \Z_2 \times \Z_2$, and is thus completely realized in the picture: it is the direct product between the rectangle isometry group $\Z_2 \times \Z_2$ and the self-polarity group $\Z_2$. This example answers positively B.~Gr\"unbaum's problem~\cite[p.\,169, Question~5]{Grunbaum1}:

\begin{fact}
There exist topological \conf{19}{4}s realized with non-trivial isomorphism groups.
\end{fact}

\begin{figure}[h]
  \centerline{\includegraphics[scale=.7]{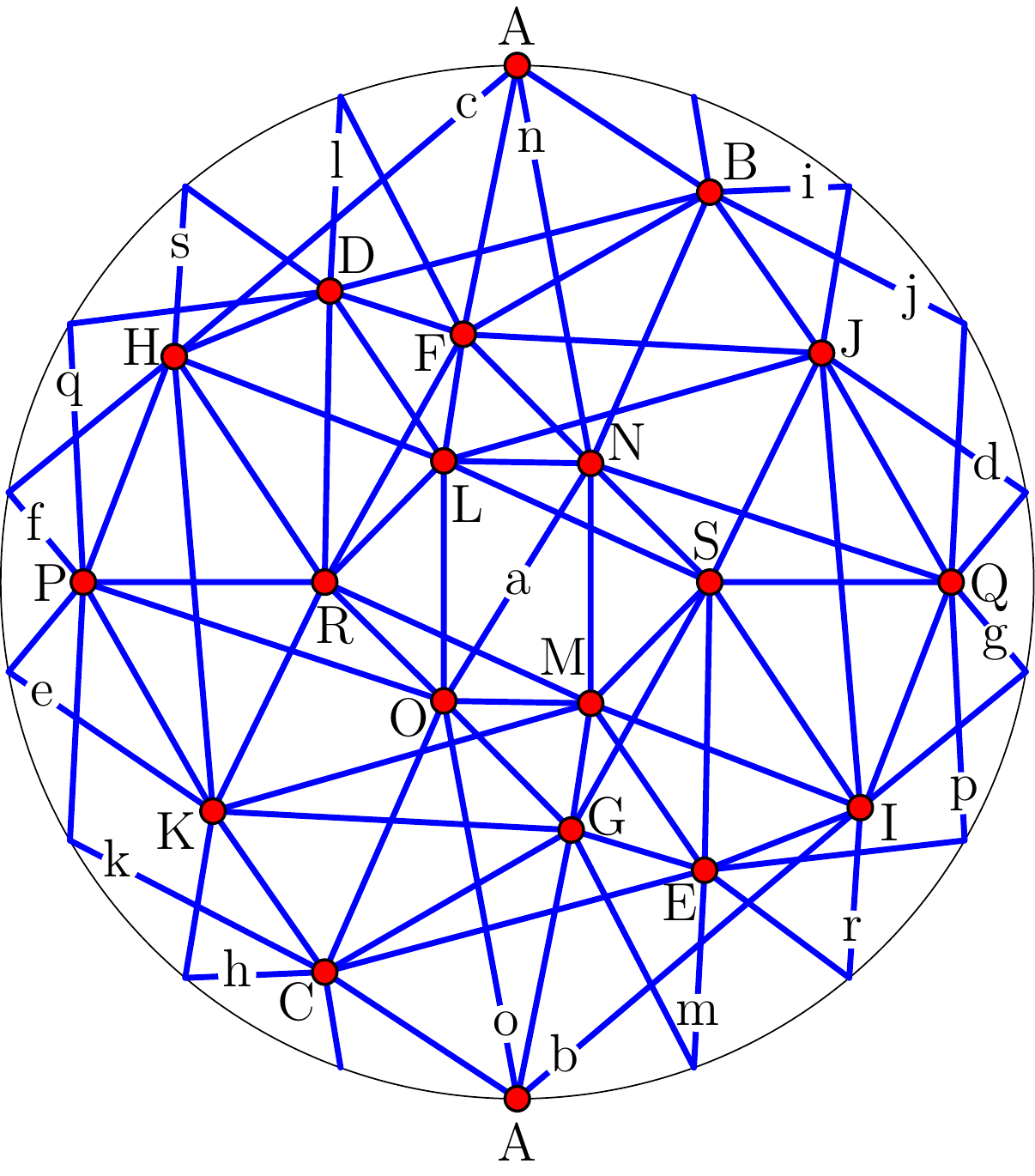}\quad\includegraphics[scale=.7]{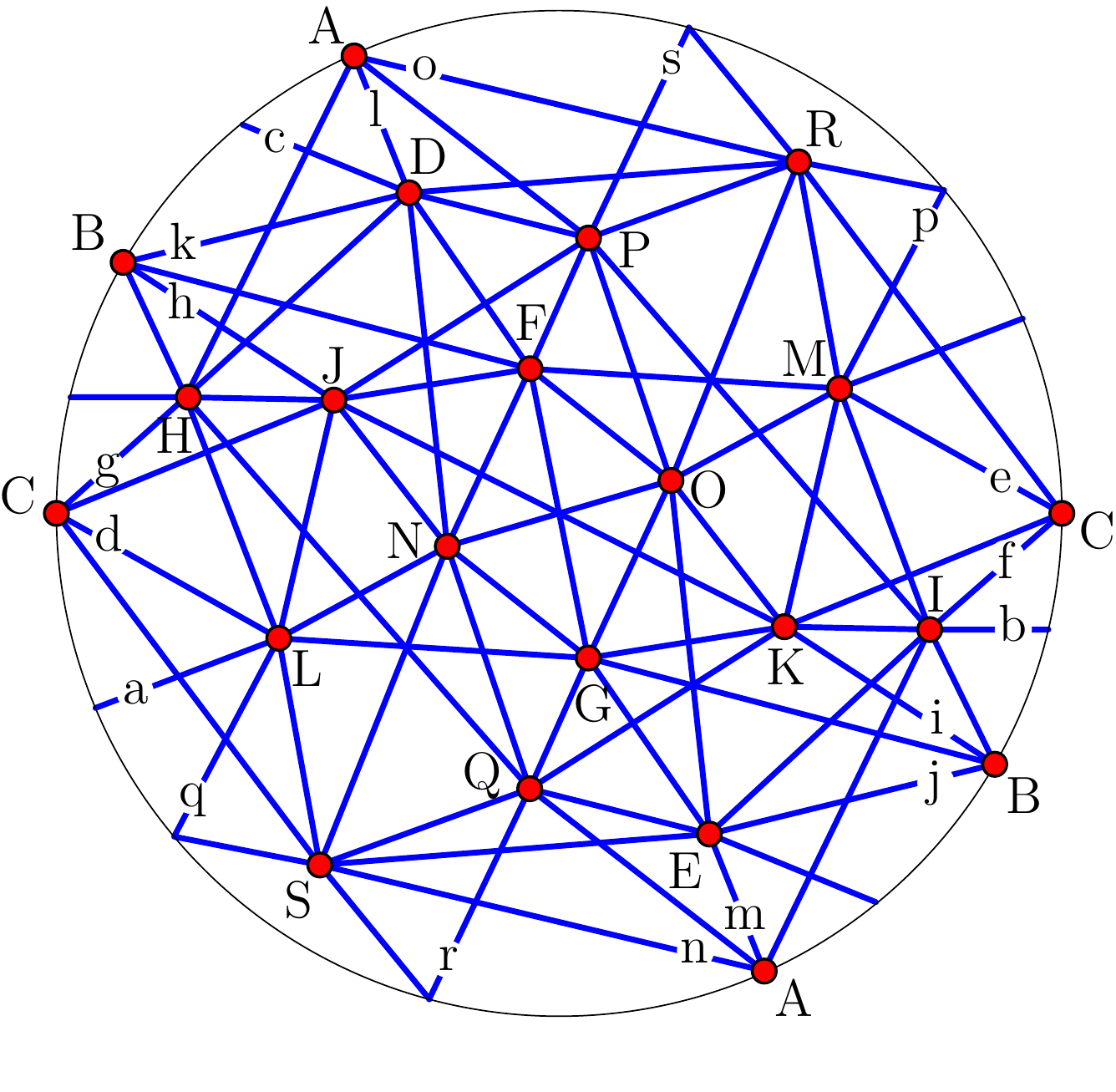}}
  \vspace*{-.2cm}
  \caption{Two centrally symmetric topological \conf{19}{4}s.}
  \label{fig:configuration_19_4_2-3}
\end{figure}

\begin{figure}[h]
  \centerline{\includegraphics[scale=.7]{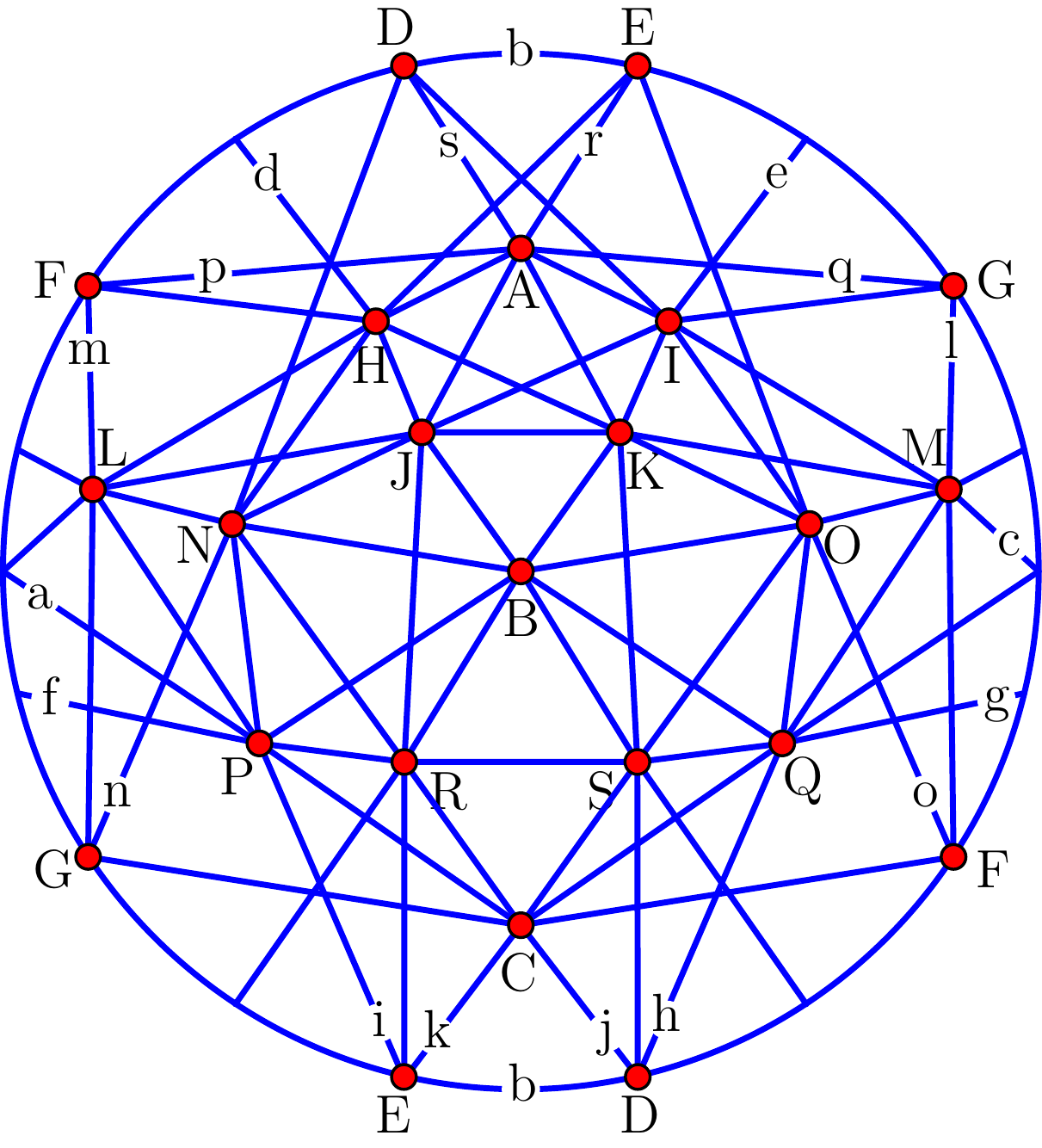} \quad \includegraphics[scale=.7]{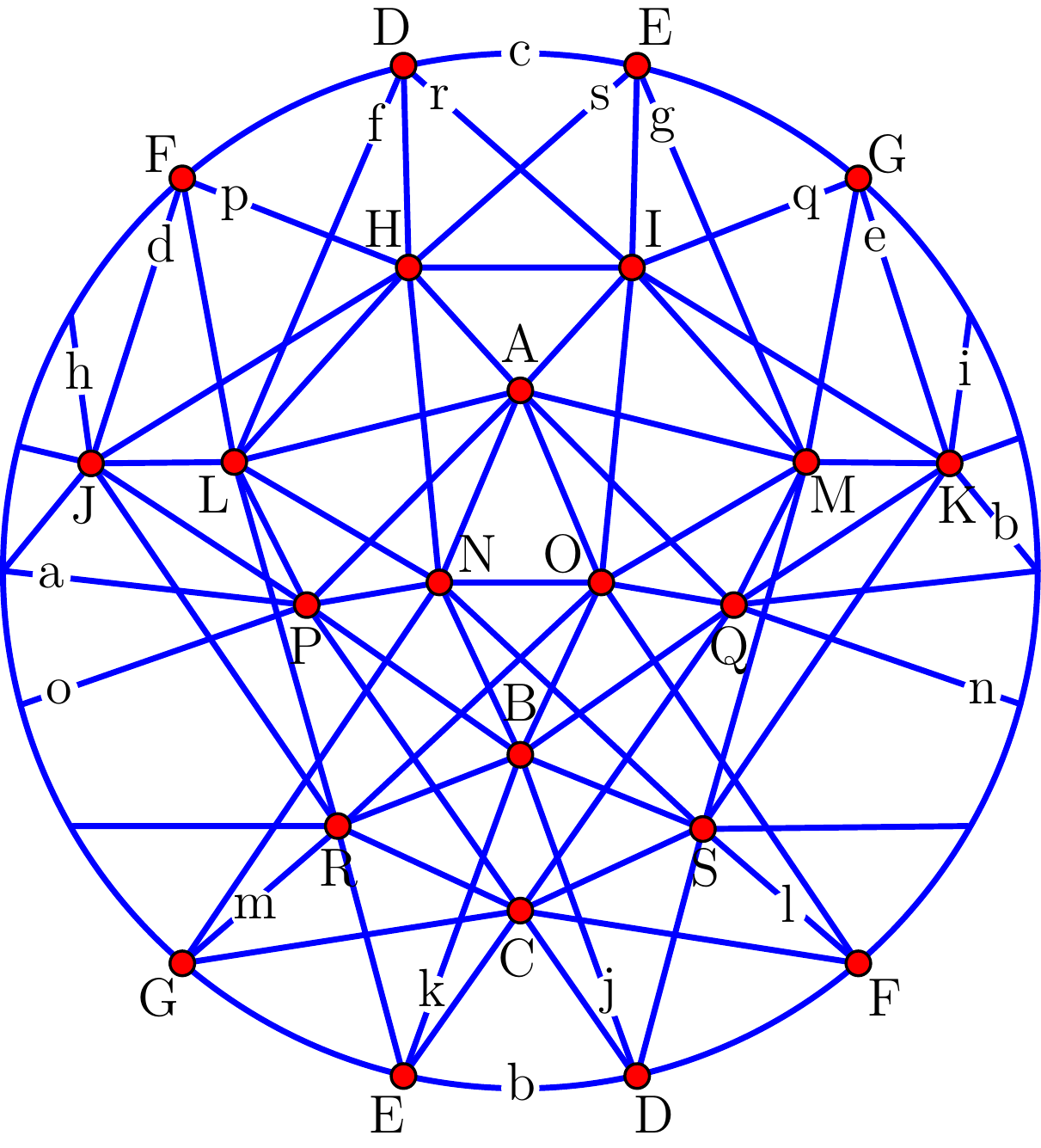}}
  \vspace*{-.2cm}
  \caption{Two topological \conf{19}{4}s with vertical symmetry.}
  \label{fig:configuration_19_4_4-5}
\end{figure}

Besides the configuration of Figure~\ref{fig:configuration_19_4_1}, there are four other topological \conf{19}{4}s with combinatorial automorphism group of order~$8$. They are represented in Figures~\ref{fig:configuration_19_4_2-3} and~\ref{fig:configuration_19_4_4-5}. We have labeled the points and lines of these configurations such that the non-trivial isometry is given by a nice permutation and the self-duality by the permutation (A,a)(B,b)\,\dots\,(R,r)(S,s). 
We detail below for each configuration its \pl{} incidences, and a generating system of its combinatorial automorphism group.

Observe that for these four configurations, we did not manage to obtain a representation where the full combinatorial automorphism group acts as isometries or polarities. It is not surprising for the configurations whose automorphism group is the dihedral group~$D_8$ since it is already impossible to construct a topological \conf{19}{4} with a $C_4$ symmetry. Indeed, since $19$ is odd, an automorphism of order~$4$ should fix a line~$\ell$ and rotate the four segments of~$\ell$ delimited by the points of the configuration. This is impossible since these four segments cross $6$ lines of the configuration, and~$6$ is not a multiple of~$4$.

\newpage

\newpage

\centerline{\textbf{\conf{19}{4} in Figure~\ref{fig:configuration_19_4_2-3}\,(left)}}

\medskip
\centerline{
\begin{tabular}{c|cccccccccccccccccccc}
  lines 		& a & b & c & d & e & f & g & h & i & j & k & l & m & n & o & p & q & r & s \\
  \hline
  			& B & A & A & J & K & I & H & C & B & B & C & D & E & A & A & D & E & D & E \\
  points 	& C & B & C & L & M & M & L & G & F & D & E & G & F & G & F & F & G & I & H \\
  in lines 	& O & J & K & R & S & O & N & S & R & H & I & O & N & M & L & J & K & S & R \\
    			& N & I & H & P & Q & P & Q & J & K & P & Q & R & S & N & O & Q & P & L & M
\end{tabular}
}
% [[1,2,14,13],[0,1,9,8],[0,2,10,7],[9,11,17,15],[10,12,18,16],[8,12,14,15],[7,11,13,16],[2,6,18,9],[1,5,17,10],[1,3,7,15],[2,4,8,16],[3,6,14,17],[4,5,13,18],[0,6,12,13],[0,5,11,14],[3,5,9,16],[4,6,10,15],[3,8,18,11],[4,7,17,12]]

\medskip
\noindent Automorphism group generated by: 
\begin{center}
(A)(B,C)(D,E)(F,G)(H,I)(J,K)(L,M)(N,O)(P,Q)(R,S) \\
(A)(B,O)(C,N)(D)(E)(F,I)(G,H)(J,L)(K,M)(P,R)(Q,S) 
\end{center}
together with the self-duality (A,a)(B,b)\,\dots\,(R,r)(S,s), which is not a self-polarity.

\ssep

\centerline{\textbf{\conf{19}{4} in Figure~\ref{fig:configuration_19_4_2-3}\,(right)}}

\medskip
\centerline{
\begin{tabular}{c|cccccccccccccccccccc}
  lines 		& a & b & c & d & e & f & g & h & i & j & k & l & m & n & o & p & q & r & s \\
  \hline
  			& L & H & D & C & C & C & C & B & B & B & B & A & A & A & A & K & J & G & F \\
  points 	& M & I & E & K & J & I & H & G & F & E & D & D & E & H & I & M & L & O & N \\
  in lines 	& O & K & G & G & F & E & D & N & O & Q & P & N & O & L & M & S & R & R & S \\
    			& N & J & F & L & M & S & R & J & K & H & I & Q & P & S & R & Q & P & Q & P
\end{tabular}
}
% [[11,12,14,13],[7,8,10,9],[3,4,6,5],[2,10,6,11],[2,9,5,12],[2,8,4,18],[2,7,3,17],[1,6,13,9],[1,5,14,10],[1,4,16,7],[1,3,15,8],[0,3,13,16],[0,4,14,15],[0,7,11,18],[0,8,12,17],[10,12,18,16],[9,11,17,15],[6,14,17,16],[5,13,18,15]]

\medskip
\noindent Automorphism group generated by: 
\begin{center}
(A)(B)(C)(D,E)(F,G)(H,I)(J,K)(L,M)(N,O)(P,Q)(R,S) \\
(A)(B,C)(D,I,E,H)(F,K,G,J)(L,N,M,O)(P,S,Q,R)
\end{center}
together with the self-polarity (A,a)(B,b)\,\dots\,(R,r)(S,s).

\ssep

\centerline{\textbf{\conf{19}{4} in Figure~\ref{fig:configuration_19_4_4-5}\,(left)}}

\medskip
\centerline{
\begin{tabular}{c|cccccccccccccccccccc}
  lines 		& a & b & c & d & e & f & g & h & i & j & k & l & m & n & o & p & q & r & s \\
  \hline
  			& P & D & J & B & B & B & B & D & E & C & C & C & C & G & F & A & A & A & A \\
  points 	& Q & E & K & J & K & O & N & I & H & D & E & G & F & I & H & F & G & E & D \\
  in lines 	& S & G & M & H & I & M & L & O & N & N & O & M & L & J & K & M & L & R & S \\
    			& R & F & L & S & R & P & Q & Q & P & R & S & Q & P & N & O & I & H & J & K
\end{tabular}
}
% [[15,16,18,17],[3,4,6,5],[9,10,12,11],[1,9,7,18],[1,10,8,17],[1,14,12,15],[1,13,11,16],[3,8,14,16],[4,7,13,15],[2,3,13,17],[2,4,14,18],[2,6,12,16],[2,5,11,15],[6,8,9,13],[5,7,10,14],[0,5,12,8],[0,6,11,7],[0,4,17,9],[0,3,18,10]]

\noindent Automorphism group generated by: 
\begin{center}
(A)(B)(C)(D,E)(F,G)(H,I)(J,K)(L,M)(N,O)(P,Q)(R,S) \\
(A,C)(B)(D,F,E,G)(H,N,I,O)(J,Q,K,P)(L,R,M,S)
\end{center}
together with the self-duality (A,a)(B,b)\,\dots\,(R,r)(S,s), which is not a self-polarity.

\ssep

\centerline{\textbf{\conf{19}{4} in Figure~\ref{fig:configuration_19_4_4-5}\,(right)}}

\medskip
\centerline{
\begin{tabular}{c|cccccccccccccccccccc}
  lines 		& a & b & c & d & e & f & g & h & i & j & k & l & m & n & o & p & q & r & s \\
  \hline
  			& N & H & D & C & C & C & C & B & B & B & B & F & G & A & A & A & A & D & E \\
  points 	& O & I & E & F & G & D & E & P & Q & D & E & L & M & L & M & H & I & I & H \\
  in lines 	& Q & K & G & J & K & L & M & J & K & H & I & N & O & J & K & F & G & M & L \\
    			& P & J & F & R & S & P & Q & S & R & N & O & S & R & Q & P & O & N & S & R
\end{tabular}
}
% [[13,14,16,15],[7,8,10,9],[3,4,6,5],[2,5,9,17],[2,6,10,18],[2,3,11,15],[2,4,12,16],[1,15,9,18],[1,16,10,17],[1,3,7,13],[1,4,8,14],[5,11,13,18],[6,12,14,17],[0,11,9,16],[0,12,10,15],[0,7,5,14],[0,8,6,13],[3,8,12,18],[4,7,11,17]]

\medskip
\noindent Automorphism group generated by: 
\begin{center}
(A)(B)(C)(D,E)(F,G)(H,I)(J,K)(L,M)(N,O)(P,Q)(R,S) \\
(A,B)(C)(D,F,E,G)(H,O,I,N)(J,Q,K,P)(L,R,M,S) 
\end{center}
together with the self-polarity (A,a)(B,b)\,\dots\,(R,r)(S,s).

%%%%%%%%%%%%%%%%%%%%%%%%%%%%%%
%%%%%%%%%%%%%%%%%%%%%%%%%%%%%%

\section{Geometric \conf{19}{4}s}
\label{sec:geometric}

In this section, we present our techniques to search for geometric realizations of the topological \conf{19}{4}s discussed in Section~\ref{sec:topological}. Observe already that we can restrict our attention to $2\,125$ topological \conf{19}{4}s, keeping only one representative in each duality class. Our main tool is the \defn{construction sequence method}, which enables us to search for subconfigurations in a configuration and to test geometric realizability of configurations. We present this method in detail below although it is a classical folklore when programing on the projective plane (it is used \eg in most dynamic geometric softwares such as \Cinderella{} or \textsc{The Geometer's Sketchpad}).

\enlargethispage{.2cm}
\paragraph{Construction sequences}
Consider the problem of searching an incidence-preserving embedding~$\phi$ of a small finite \pl{} configuration~$(P,L)$ into a large \pl{} configuration~$(\Pi,\Lambda)$. Note that this problem covers two relevant situations that we will detail later on:
\begin{enumerate}[(i)]
\item if~$(\Pi,\Lambda)$ is a finite \pl{} configuration, then we are searching for subconfigurations isomorphic to~$(P,L)$ in a configuration~$(\Pi,\Lambda)$.
\item if~$\Pi$ is the set of all points and~$\Lambda$ the set of all lines of the plane, then we are testing the geometric realizability of~$(P,L)$.
\end{enumerate}
In both situations, it is natural to start fixing the image under~$\phi$ of an arbitrary projective base~$P_1$ of~$(P,L)$, then construct the image of the set~$L_1$ of all lines of~$L$ joining two points of~$P_1$, then the image of the set~$P_2$ of all points of~$P \ssm P_1$ contained in at least two lines of~$L_1$, \etc{} If this procedure finishes, we can easily test whether the final embedding~$\phi$ indeed respects all incidences of~$(P,L)$. It might however happen that the procedure described above does not finish: at some point, it might happen that none of the remaining point (or line) is incident with two constructed lines (or points). In this case, we have to consider all possible positions for constructing a new point (or line) before getting back to the procedure.

We formalize this intuitive description as follows. We denote by~$p \vee p'$ the unique line of~$L$ passing through two points~$p$ and~$p'$ of~$P$ (if it exists). Similarly, let~$\ell \wedge \ell'$ be the unique point of~$P$ contained in two lines~$\ell$ and~$\ell'$ of~$L$ (if it exists). A \defn{projective base} of the combinatorial \pl{} configuration~$(P,L)$ is a set~$B$ of four points of~$P$ such that for every triple~$T \subset B$, there exists a line of~$L$ containing precisely two points of~$T$. This ensures that no three points of~$B$ can be aligned, even in a larger \pl{} configuration containing~$(P,L)$. A \defn{construction sequence} for~$(P,L)$ is a sequence~$(X_i)$ of subsets of~$P \sqcup L$ such that
\begin{enumerate}[(i)]
\item $X_0$ is a projective base of~$(P,L)$.
\item $X_{i+1}$ is the set of all elements of~$P \sqcup L$ not in~$\bigcup_{j \le i} X_j$ incident to at least two elements of~$\bigcup_{j \le i} X_{j}$ if it is non-empty. Otherwise, $X_{i+1}$ is a single element of~$P \sqcup L$ not in~$\bigcup_{j \le i} X_j$ incident to one element of~$\bigcup_{j \le i} X_j$. Note that~$X_i \subset P$ when $i$ is even, and $X_i \subset L$ otherwise. 
\end{enumerate}
It models the intuitive notion of sequence of construction for the configuration~$(P,L)$: once we choose the image under~$\phi$ of the projective base~$X_0$, we proceed to a sequence of construction of points and lines defining at each step the image of~$X_{i+1}$ using the image of~$\bigcup_{j \le i} X_{j}$. When~$X_{i+1}$ is formed by a single line (or point) containing only one point (or line) already constructed, the situation is underdeterminated and results either in different branches of the procedure if we search for subconfigurations, or in the introduction of a free variable if we test geometric realizability. These two situations are described separately below.

\paragraph{Subconfigurations}
Our first task is to search for subconfigurations of a finite \pl{} configuration~$(\Pi,\Lambda)$ which are isomorphic to another \pl{} configuration~$(P,L)$. Note that it can be seen as a particular instance of subgraph isomorphism, but that the construction sequence method will significantly speed up the research.

\renewcommand{\algorithmicforall}{\textbf{for each}}
\newcommand{\LINEIF}[3]{%
    \STATE\algorithmicif\ {#1}\ \algorithmicdo\ {#2} \algorithmicelse\ {#3} \algorithmicend\ \algorithmicif%
}

\begin{algorithm}[p]
\caption{--- Subconfigurations}
\label{algo:subconfigurations}
\begin{algorithmic}

\medskip
\REQUIRE Two finite \pl{} configurations $(P,L)$ and~$(\Pi,\Lambda)$.
\ENSURE  All subconfigurations of~$(\Pi,\Lambda)$ isomorphic to~$(P,L)$.

\medskip
\STATE Choose an arbitrary projective base~$(p_1,p_2,p_3,p_4)$ of~$(P,L)$.
\FORALL{projective base~$(\pi_1,\pi_2,\pi_3,\pi_4)$ of~$(\Pi,\Lambda)$}
	\STATE Initialize the images $\phi(p_i) \eqdef \pi_i$ for~$i \in [4]$ and the definition domain $D \eqdef \{p_1, p_2, p_3, p_4\}$.
	\REPEAT
		\STATE Initialize a boolean $\textsc{ncl} \eqdef \textrm{false}$ (witnessing whether we found new constructible lines).
		\FORALL{line~$\ell \in L \ssm D$}
			\IF{there exists points $p,p' \in \ell \cap D$}
				\STATE Set~$\phi(\ell) \eqdef \phi(p) \vee \phi(p')$ and update $D \eqdef D \cup \{\ell\}$ and $\textsc{ncl} \eqdef \textrm{true}$.
				\STATE Check that~$q \in \ell \iff \phi(q) \in \phi(\ell)$ for all~$q \in P \cap D$. Otherwise reject.
			\ENDIF
		\ENDFOR
		\IF{\textsc{ncl}}
			\STATE Dualize $P \leftrightarrow L$ and $\Pi \leftrightarrow \Lambda$ and repeat.
		\ELSE
			\STATE Choose an arbitrary line~$\ell \in L \ssm D$ such that there is one point~$p \in \ell \cap D$.
			\FORALL{line~$\lambda \in \Lambda \ssm \phi(D)$ containing~$\phi(p)$}
				\STATE Set $\phi(\ell) \eqdef \lambda$ and update $D \eqdef D \cup \{\ell\}$.
				\STATE Check that~$q \in \ell \iff \phi(q) \in \phi(\ell)$ for all~$q \in P \cap D$. Otherwise reject.
				\STATE Dualize $P \leftrightarrow L$ and $\Pi \leftrightarrow \Lambda$ and repeat.
			\ENDFOR
		\ENDIF
	\UNTIL{$D = P \sqcup L$.}
	\RETURN $\phi$.
\ENDFOR
\end{algorithmic}
\bigskip
\end{algorithm}

\begin{figure}[p]
  \vspace{.5cm}
  \centerline{\includegraphics[scale=.65]{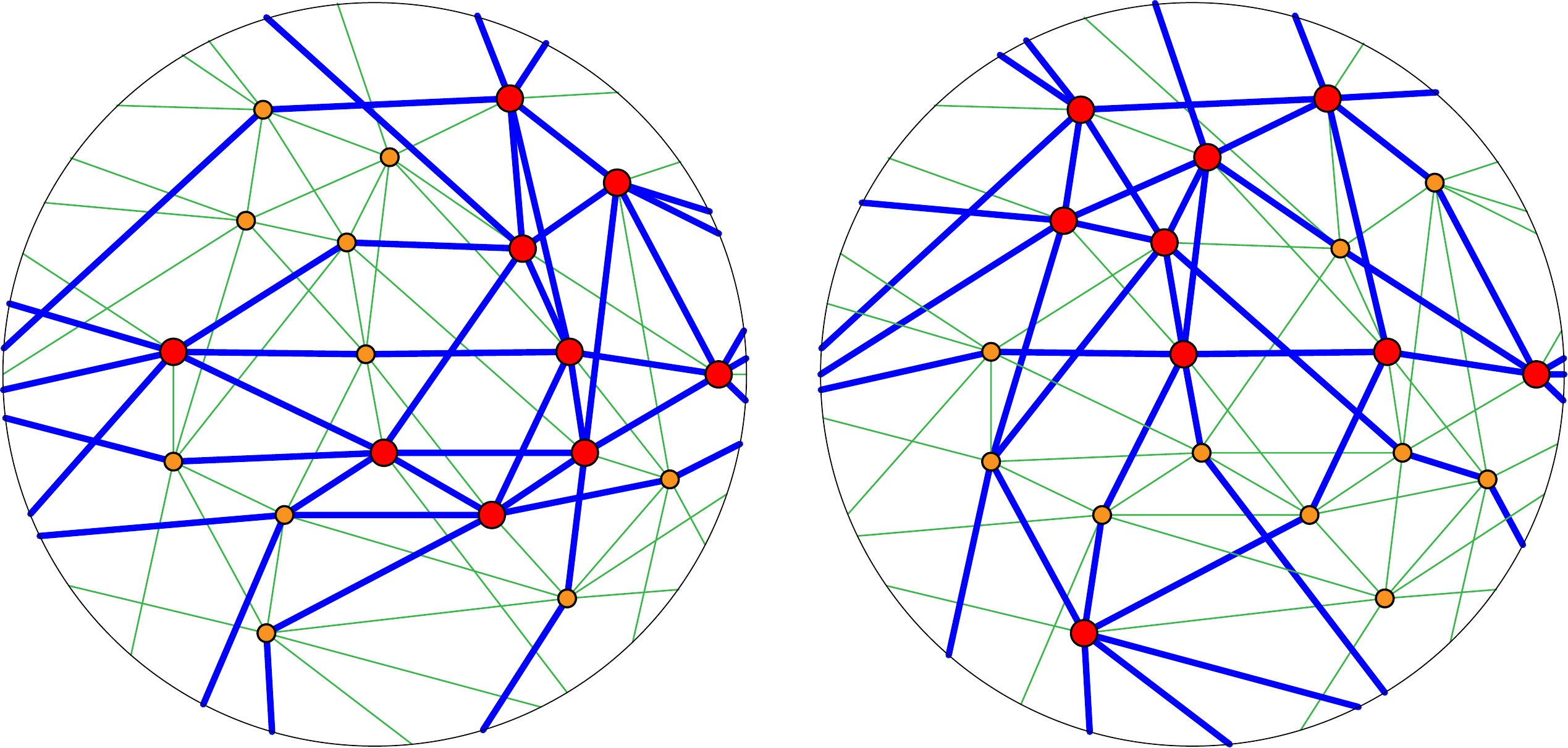}}
  \caption{A topological \conf{19}{4} containing a Pappus (left) and a non-Pappus (right) subconfiguration, whose points and lines are bolded.}
  \label{fig:configuration_19_4_Pappus}
\end{figure}

Algorithm~\ref{algo:subconfigurations} exploits construction sequences to search for subconfigurations of a finite \pl{} configuration. See page~\pageref{algo:subconfigurations}. The progress of this algorithm depends on the choice of the initial projective base: it might create several branches for certain choices of this base, and much less for other bases. Similarly, in case of branches, the choice of the line~$\ell \in L \ssm D$ can also affect the number of further branches needed to complete the construction sequence. This can be easily optimized over all possible construction sequences. In any case, whatever choices are made, the algorithm will always end up with all subconfigurations isomorphic to~$(P,L)$ in~$(\Pi,\Lambda)$.

In our presentation of this algorithm, we are always constructing lines but we dualize at each step of the algorithm, thus inverting the role of points and lines at each step. We could have instead written twice the same code, exchanging points with lines in the second copy. We have preferred this version to shorten the presentation.

We use subconfigurations to test for example Pappus' theorem in our configurations. For that, it suffices to search in each \conf{19}{4} for the non-Pappus configuration --- obtained from Pappus configuration represented in Figure~\ref{fig:smallConfs}\,(right) by deleting a single incidence. We illustrate examples of Pappus and non-Pappus subconfigurations of a \conf{19}{4} in Figure~\ref{fig:configuration_19_4_Pappus}. Using Algorithm~\ref{algo:subconfigurations}, we could test efficiently Pappus' and Desargues' theorems in our $2\,125$ topological \conf{19}{4}s (one per duality class), and we obtain the following result.

\begin{fact}
Among the $2\,125$ topological \conf{19}{4}s (up to combinatorial equivalence and duality), only $512$ configurations are compatible with both Pappus' and Desargues' theorems.
\end{fact}

\paragraph{Geometric realizability}
Our second important task is to test whether a combinatorial \pl{} configuration~$(P,L)$ can be realized geometrically in the projective plane. Observe first that it is clearly an instance of the \defn{Existential Theory of the Reals} (ETR): it can be expressed as a system of polynomial equalities~$\bE$ and inequalities~$\bI$ on a set~$\Theta$ of real variables. A naive approach consists in assigning two variables to each point of~$P$ and to each line of~$L$, and to construct quadratic equalities and inequalities according to the \pl{} incidences (one quadratic equality per incidence, and one quadratic inequality per missing incidence).

\begin{remark}
Working in the projective plane, we represent both points and lines of~$\bP$ as (pairs of antipodal) vectors. In other words, a line of~$\bP$ is represented by its normal direction. All geometric primitives on points and lines in~$\bP$ then correspond to simple computations on their representing vectors:
\begin{enumerate}[(i)]
\item a point is incident to a line iff their representing vectors are orthogonal, and
\item the vector representing the point defined by two lines is the cross product of the vectors representing these two lines. Similarly, the vector representing the line defined by two points is the cross product of the vectors representing these two points.
\end{enumerate}
\end{remark}

Algorithm~\ref{algo:realization} exploits construction sequences to test the geometric realizability of a combinatorial \pl{} configuration~$(P,L)$. See page~\pageref{algo:realization}. It still expresses the problem of geometric realizability of a configuration as an instance of ETR. However, its contribution is to reduce drastically the number of variables needed, to the price of increasing substantially the degree of the polynomials involved in the equalities and inequalitites.

\begin{algorithm}
\caption{--- Geometric realizability}
\label{algo:realization}
\begin{algorithmic}

\medskip
\REQUIRE A finite \pl{} configuration $(P,L)$.
\ENSURE  Decides whether $(P,L)$ is geometrically realizable in the projective plane~$\bP$.

\medskip
\STATE Choose arbitrary projective bases~$(p_1,p_2,p_3,p_4)$ of~$(P,L)$ and~$(\pi_1,\pi_2,\pi_3,\pi_4)$ of~$\bP$.
\STATE Initialize the images $\phi(p_i) \eqdef \pi_i$ for~$i \in [4]$ and the definition domain $D \eqdef \{p_1, p_2, p_3, p_4\}$.
\STATE Initialize the collections of equalities~$\bE \eqdef \varnothing$, of inequalities~$\bI \eqdef \varnothing$, and of variables~$\Theta \eqdef \varnothing$.
\REPEAT
	\STATE Initialize a boolean $\textsc{ncl} \eqdef \textrm{false}$ (witnessing whether we found new constructible lines).
	\FORALL{line~$\ell \in L \ssm D$}
		\IF{there exists points $p,p' \in \ell \cap D$}
			\STATE Set~$\phi(\ell) \eqdef \phi(p) \vee \phi(p')$ and update $D \eqdef D \cup \{\ell\}$ and $\textsc{ncl} \eqdef \textrm{true}$.
			\FORALL{$q \in P \cap D$}
				\LINEIF{$q \in \ell$}{$\bE \eqdef \bE \cup \{\phi(q) \cdot \phi(\ell) = 0\}$}{$\bI \eqdef \bI \cup \{\phi(q) \cdot \phi(\ell) \ne 0\}$}.
			\ENDFOR
		\ENDIF
	\ENDFOR
	\IF{\NOT\textsc{ncl}}
		\STATE Choose an arbitrary line~$\ell \in L \ssm D$ such that there is one point~$p \in \ell \cap D$.
		\STATE Introduce a variable~$\theta \in \R$ and update~$\Theta \eqdef \Theta \cup \{\theta\}$.
		\STATE Define a parametrization~$\theta \in \R \mapsto \lambda_\theta$ of the lines passing through~$\phi(p)$.
		\STATE Set $\phi(\ell) \eqdef \lambda_\theta$ and update $D \eqdef D \cup \{\ell\}$.
		\FORALL{$q \in P \cap D$}
			\LINEIF{$q \in \ell$}{$\bE \eqdef \bE \cup \{\phi(q) \cdot \phi(\ell) = 0\}$}{$\bI \eqdef \bI \cup \{\phi(q) \cdot \phi(\ell) \ne 0\}$}.
		\ENDFOR
	\ENDIF
	\STATE Dualize $P \leftrightarrow L$ and $\Pi \leftrightarrow \Lambda$ and repeat.
\UNTIL{$D = P \sqcup L$.}
\IF{the system of equalities~$\bE$ and inequalities~$\bI$ in the variables~$\Theta$ has a solution~$\Theta_\circ$}
	\STATE Replace~$\Theta$ by~$\Theta_\circ$ in~$\phi$.
	\RETURN $\phi$.
\ELSE
	\STATE Reject.
\ENDIF
\end{algorithmic}
\bigskip
\end{algorithm}

Observe that in this algorithm, it is sufficient to consider only one arbitrary projective base of~$\bP$ since they are all projectively equivalent. Observe also that we have chosen again to shorten the code by dualizing the configuration at each step of the algorithm.

To solve the ETR instance~$(\bE, \bI, \Theta)$, we use the computer algebra system \Maple{}. Although the resulting system contains a priori equalities and inequalities of high degree in several variables, we observed that among the $512$ topological \conf{19}{4}s compatible with Pappus' theorem:
\begin{enumerate}[(i)]
\item $10$ configurations admit a complete construction sequence which never introduces any variable, but do not fulfill all required incidences. They are immediately discarded.

\begin{example}
The configuration given by the incidence table
%
%[[[8,9,10,15],[3,4,5,15],[0,3,6,10],[4,6,8,18],[15,16,17,18],[0,1,2,15],[11,12,13,16],[0,7,9,16],[1,4,10,16],[1,6,9,13],[1,3,8,12],[0,11,14,17],[6,7,12,17],[2,5,13,17],[2,4,7,14],[5,9,12,14],[2,3,11,18],[10,13,14,18],[5,7,8,11]],[[9,11],[6,10],[8],[5],[17,18],[7,14,15],[12,13,16,19]],[[4,5,7,9,11,16],[2],[1,10],[17],[19],[3,14],[6,8,12,13,15,18]]]
%
\begin{center}
\begin{tabular}{c|cccccccccccccccccccc}
  lines 		& a & b & c & d & e & f & g & h & i & j & k & l & m & n & o & p & q & r & s \\
  \hline
  			& C & F & F & B & B & B & C & H & A & A & A & G & G & G & L & A & E & E & D \\
  points 	& F & I & N & C & D & N & D & M & D & H & C & L & K & J & O & B & G & L & E \\
  in lines 	& H & J & O & K & I & P & J & O & K & J & I & Q & M & N & P & E & H & M & Q \\
    			& L & K & Q & Q & O & S & M & S & S & P & R & S & P & R & R & F & I & N & R
\end{tabular}
\end{center}
\medskip
admits the complete construction sequence \\
\centerline{ABCD - degikp - IK - b - FJ - aj - H - q - E - s - QR - cn - GNO - fhlmor - LMPS,\hspace*{1cm}}
which does not introduce any variable point or line.
The contradiction in the construction sequence arises when we construct the lines h and o (undesired incidences R-h and H-o are forced) and when we then construct the last points L, M, P, and S (desired incidences M-m, P-o, M-r, L-r, P-m, S-i, L-o, S-l are missing, and undesired incidences S-o, M-o are forced). Although they are computed by our \Maple{} code, the reader can check these additional and missing incidences by performing the given construction sequence with a dynamic geometry software like \Cinderella{}.
\end{example}

\enlargethispage{.3cm}
\item $486$ configurations admit a construction sequence which results in an ETR instance involving only one variable, and for which the equalities and inequalities of degree at most four already produce a contradiction (note that such a system can be solved by radicals). In fact, among these $496$ cases, most of them already contain a contradiction in their equalities and inequalities of smaller degree. The following table shows the repartition (and percentage) of the minimal degree leading to a contradiction in the $496$ construction sequences that we have considered.
\begin{center}
\renewcommand{\arraystretch}{1.2}
\begin{tabular}{c|cccccc}
degree & $0$ & $1$ & $2$ & $3$ & $4$ & \\
\hline
number of cases & $15$ & $92$ & $192$ & $132$ & $55$ \\
\hline
proportion (\%) & $3$ & $19$ & $40$ & $27$ & $11$
\end{tabular}
\end{center}
Note that each configuration could admit another construction sequence which yields a contradiction of smaller degree. We did not try to optimize further than getting, for each configuration, a construction sequence leading to a contradiction of degree at most four.

\begin{example}
The configuration given by the incidence table
%
%([[15,16,17,18],[8,9,10,15],[3,7,10,16],[3,8,13,17],[3,4,5,15],[0,1,2,15],[12,13,14,16],[0,4,6,16],[3,9,11,14],[1,6,10,14],[0,7,11,17],[2,5,14,17],[1,4,11,13],[0,8,12,18],[1,7,9,18],[2,6,13,18],[2,4,9,12],[5,10,11,12],[5,6,7,8]],[[5],[12,18,19],[11],[9],[7,10,15],[6,8,13,14,16,17]],[[4,9,11,16,17,18],[6],[8],[12],[10,15],[2,7,13,14,19],[1,3,5]])
%
\begin{center}
\begin{tabular}{c|cccccccccccccccccccc}
  lines 		& a & b & c & d & e & f & g & h & i & j & k & l & m & n & o & p & q & r & s \\
  \hline
  			& F & F & F & C & E & E & H & C & B & B & B & I & G & D & G & A & A & A & A \\
  points 	& H & J & L & D & H & L & J & K & D & I & C & K & N & G & I & B & C & D & N \\
  in lines 	& K & M & P & E & M & R & P & O & N & O & J & M & Q & M & J & E & G & K & O \\
    			& N & O & Q & I & Q & S & S & S & S & Q & R & R & R & P & L & F & H & L & P
\end{tabular}
\end{center}
\medskip
admits the construction sequence \\
\centerline{ABCD - dikpqr - E - \boxed{\textrm{f}} - LRS - h - K - l - I - jo - GJO - bgmns - FHMNPQ - ace,\hspace*{1cm}}
which introduces only one variable when constructing line~f (boxed in the sequence). Let us follow the beginning of the construction sequence. First, we send the projective base~$ABCD$ of the configuration to the projective base~$\phi(A) = [1, 0, 0]$, $\phi(B) = [0, 1, 0]$, $\phi(C) = [0, 0, 1]$, and $\phi(D) = [1, 1, 1]$ of the projective plane~$\bP$. We can then construct the images of the line $\phi(d) = \phi(C \vee D) = \phi(C) \vee \phi(D) = [-1, 1, 0]$, and similarly $\phi(i) = [1, 0, -1]$, $\phi(k) = [1, 0, 0]$, $\phi(p) = [0, 0, 1]$, $\phi(q) = [0, -1, 0]$, $\phi(r) = [0, -1, 1]$. In turn, we obtain the image of the point~$\phi(E) = \phi(d \wedge p) = \phi(d) \wedge \phi(p) = [1, 1, 0]$. At that stage, the construction sequence is blocked since there is no more element of the configuration incident to two elements whose images are already determined. We therefore decide to construct the line~$\phi(f) = [1, -1, \theta]$ containing the point~$\phi(E)$ and parametrized by the variable~$\theta \in \R$. We can then start again the construction using this last constructed line~$\phi(f)$. We construct the points $\phi(L) = [\theta-1, -1, -1]$, $\phi(R) = [0, \theta, 1]$, and $\phi(S) = [1, \theta+1, 1]$. The incidences of these points with the lines already constructed force $\theta \notin \{0,-1,1\}$. While we keep running the construction sequence, we do not need any further variable, but we obtain many more conditions on~$\theta$ (our \Maple{} code produces $9$ equalities and $113$ inequalities). One of the inequalities simplifies to~$0 \ne 0$, meaning that whatever the parameter~$\theta$ is, an undesired incidence is forced. Even without considering any equality and inequality involving the variable~$\theta$, we thus conclude that the system has no solution.
\end{example}

\item the remaining $16$ configurations require two variables. These last cases are a bit more complicated to handle, and we therefore start with two examples.
\begin{example}
\label{exm:2variables}
The configuration given by the incidence table
%
% [[[15,16,17,18],[8,9,10,15],[0,3,6,10],[0,7,9,11],[3,4,5,15],[0,1,2,15],[0,13,14,16],[7,8,12,16],[4,6,11,16],[3,9,12,14],[1,5,11,14],[1,4,8,13],[1,3,7,17],[6,9,13,17],[2,10,14,17],[2,5,12,13],[2,4,7,18],[5,6,8,18],[10,11,12,18]],[[6],[5],[10,13],[14,15],[7,11,17],[8,9,12,16,18,19]],[[1,10,11,16],[2],[4],[8,18],[3,7,15],[5,6,12,14,17,19],[9,13]]]
%
\begin{center}
\begin{tabular}{c|cccccccccccccccccccc}
  lines 		& a & b & c & d & e & f & g & h & i & j & k & l & m & n & o & p & q & r & s \\
  \hline
  			& C & F & F & C & E & E & C & D & B & B & B & D & H & G & G & A & A & A & A \\
  points 	& D & K & O & E & I & K & I & H & H & D & C & I & J & L & J & B & G & M & Q \\
  in lines 	& F & L & P & J & L & P & N & M & L & J & O & K & P & N & K & E & H & N & R \\
    			& G & M & Q & M & Q & R & R & Q & R & N & S & S & S & P & O & F & I & O & S
\end{tabular}
\end{center}
\medskip
admits the construction sequence \\
\centerline{ABCD - ajkp - F - \boxed{\textrm{b}} - \boxed{\textrm{E}} - d - JM - hr - NO - cgo - GKQ - eflnqs - HILPRS - im,\hspace*{1cm}}
which introduces only two free variables when constructing line~b and point~E (boxed in the sequence). As earlier, let us perform the first steps of the construction sequence. Starting from the projective base~$\phi(A) = [1, 0, 0]$, $\phi(B) = [0, 1, 0]$, $\phi(C) = [0, 0, 1]$, and $\phi(D) = [1, 1, 1]$, we construct the lines $\phi(a) = [-1, 1, 0]$, $\phi(j) = [1, 0, -1]$, $\phi(k) = [1, 0, 0]$, $\phi(p) = [0, 0, 1]$, and then the point $\phi(F) = [1, 1, 0]$. We then need to introduce variables~$\theta$ and~$\vartheta$ in the next two steps, first for the line $\phi(b) = [1, -1, \theta]$ and then for the next point $\phi(E) = [1, \vartheta, 0]$. Observe that we immediately obtain that $\theta \ne 0$ since $C \notin b$, and that $\vartheta - 1 \ne 0$ since $E \notin a$. We can then construct the line~$\phi(d) = [-\vartheta, 1, 0]$ and obtain that~$\vartheta \ne 0$ since~$A \notin d$. While we keep running the construction sequence, we do not need any further variables, but we obtain many conditions on~$\theta$ and~$\vartheta$ (our \Maple{} code produces $9$ equalities and $166$ inequalities). Among them, the incidence between point I and line q leads to the equation $\theta^2\vartheta(\vartheta-1)^3 = 0$, which is already impossible since~$\theta \ne 0$, $\vartheta - 1 \ne 0$, and~$\vartheta \ne 0$.
\end{example}
\begin{example}
\label{exm:2variablesBis}
The configuration given by the incidence table
%
% [[[15,16,17,18],[8,9,10,15],[3,6,7,10],[0,6,12,16],[3,4,5,15],[0,1,2,15],[0,4,9,11],[1,7,11,16],[2,5,10,16],[0,3,8,13],[1,4,12,13],[1,3,14,17],[4,6,8,17],[2,9,13,17],[2,6,14,18],[7,9,12,18],[5,11,13,18],[5,7,8,14],[10,11,12,14]],[[9],[6],[15],[5],[10,17],[13,14,18],[7,8,11,12,16,19]],[[7,11,16,17],[1],[3],[19],[4,6],[9,14],[5,8,10,15,18],[2,12,13]]]
%
\begin{center}
\begin{tabular}{c|cccccccccccccccccccc}
  lines 		& a & b & c & d & e & f & g & h & i & j & k & l & m & n & o & p & q & r & s \\
  \hline
			& D & F & F & C & E & E & C & C & B & B & B & G & D & J & L & A & A & A & A \\
  points 	& F & H & I & E & G & I & D & H & J & G & C & H & K & K & O & B & D & L & O \\
  in lines 	& G & K & N & J & K & Q & M & P & M & N & I & Q & P & N & R & E & H & M & P \\
			& J & L & O & L & M & R & O & R & R & P & S & S & S & Q & S & F & I & N & Q
\end{tabular}
\end{center}
\medskip
admits the construction sequence \\
\centerline{ABCD - gkpq - I - \boxed{\textrm{a}} - F - c - O - s - \boxed{\textrm{E}} - df - JQ - in - MNR - ehjor - GHKLPS - blm,\hspace*{1cm}}
which introduces only two free variables when constructing line~a and point~E (boxed in the sequence). As earlier, let us perform the first steps of the construction sequence. Starting from the projective base~$\phi(A) = [1, 0, 0]$, $\phi(B) = [0, 1, 0]$, $\phi(C) = [0, 0, 1]$, $\phi(D) = [1, 1, 1]$, we construct the lines~$\phi(g) = [-1, 1, 0]$, $\phi(k) = [1, 0, 0]$, $\phi(p) = [0, 0, 1]$, $\phi(q) = [0, -1, 1]$, and the point~$\phi(I) = [0, -1, -1]$. We then need to introduce a free variable~$\theta$ for the line~${\phi(a) = [1, -\theta-1, \theta]}$. From this line, we construct the point~${\phi(F) = [-1-\theta, -1, 0]}$, the line~${\phi(c) = [1, -1-\theta, 1+\theta]}$, the point~${\phi(O) = [-1-\theta, -1-\theta, -\theta]}$ and then the line ${\phi(s) = [0, \theta, -1-\theta]}$. Again, we introduce a new variable~$\vartheta$ for the point ${\phi(E) = [1, \vartheta, 0]}$. From this point, we construct the lines~${\phi(d) = [-\vartheta, 1, 0]}$ and~${\phi(f) = [-\vartheta, 1, -1]}$, then the points~$\phi(J) = [-\theta, -\theta\vartheta, 1-\vartheta-\theta\vartheta]$ and~$\phi(Q) = [-1, -\theta\vartheta - \vartheta, -\theta\vartheta]$, and so on until we finally construct the line~$\phi(m)$. Along the remaining construction sequence, we do not need any further variable, but we obtain many conditions on~$\theta$ and~$\vartheta$ (our \Maple{} code produces~$8$ equalities and~$149$ inequalities). Among them, we obtain~$\theta\vartheta + \vartheta - 1 \ne 0$ since~$J \notin f$, while $(\theta\vartheta + \vartheta - 1) \big( (2 \theta^2 + 3 \theta + 1) \vartheta^2 - (2 \theta^2 + 3 \theta + 1) \vartheta + \theta \big) = 0$ since~$L \in r$ and $(\theta\vartheta + \vartheta - 1) \big( (3 \theta^3 + 3 \theta^2 - \theta - 1) \vartheta^2 - (2 \theta^3 + 5 \theta^2 - 2) \vartheta +  (\theta^2 + \theta - 1) \big) = 0$ since~$G \in j$. We therefore obtain two equations of degree~$2$ in~$\vartheta$, from which we can eliminate
\[
\vartheta = \frac{\theta^3 - 3 \theta^2 + 1}{(2 \theta + 1) (\theta^3 - 2 \theta^2 - \theta + 3)}.
\]
Plugging in this value of~$\vartheta$, we obtain a system of polynomial equalities and inequalities involving only one variable~$\theta$. We can again solve the subsystem of equations of degree at most~$4$ in~$\theta$ and check that none of the resulting solutions yields a solution for the initial system, which shows that this configuration is not geometrically realizable.
%
%[[P,Q,R,S],[I,J,K,P],[D,G,H,K],[A,G,M,Q],[D,E,F,P],[A,B,C,P],[A,E,J,L],[B,H,L,Q],[C,F,K,Q],[A,D,I,N],[B,E,M,N],[B,D,O,R],[E,G,I,R],[C,J,N,R],[C,G,O,S],[H,J,M,S],[F,L,N,S],[F,H,I,O],[K,L,M,O]]
%
%[[P,I,D,A,D,A,A,B,C,A,B,B,E,C,C,H,F,F,K],[Q,J,G,G,E,B,E,H,F,D,E,D,G,J,G,J,L,H,L],[R,K,H,M,F,C,J,L,K,I,M,O,I,N,O,M,N,I,M],[S,P,K,Q,P,P,L,Q,Q,N,N,R,R,R,S,S,S,O,O]]
%
%
\end{example}
\noindent Among the $16$ configurations involving two variables, 
\begin{itemize}
\item $12$ cases can be handled as in Example~\ref{exm:2variables}. Namely, at least one equality of~$\bE$ factors into smaller polynomials which all appear as factors of at least one inequality of~$\bI$, thus providing a simple contradiction, although the construction sequence involves two variables.
\item the remaining $4$ cases can be handled as in Example~\ref{exm:2variablesBis}. Namely, after simplification by all factors of the inequalities of~$\bI$, there are always two equalities of~$\bE$ of degree two in one of the variables. We can therefore eliminate this variable to obtain a simplified system of equalities and inequalities involving a single variable. This system can be proved to have no solution by considering only equalities of degree at most~$4$ and proving that the resulting solutions do not yield solutions of the initial system.
\end{itemize}
In particular, even for the cases involving two variables, we did not need any sophisticated techniques (for example based on Gr\"obner bases) to ensure non feasibility of all these instances of ETR. All computations were handled with the computer algebra system \Maple{}.
\end{enumerate}

\medskip
This concludes our study of geometric \conf{19}{4}s, and leads to the following surprising statement, which closes one of the last remaining cases in the quest for \conf{n}{4}s.

\begin{fact}
There is no geometric \conf{19}{4}.
\end{fact}

%%%%%%%%%%%%%%%%%%%%%%%%%%%%%%
%%%%%%%%%%%%%%%%%%%%%%%%%%%%%%

\newpage
\bibliographystyle{alpha}
\bibliography{194Configurations.bib}

\begin{thebibliography}{BGS09}

\bibitem[BGS09]{BokowskiGrunbaumSchewe}
J{\"u}rgen Bokowski, Branko Gr{\"u}nbaum, and Lars Schewe.
\newblock Topological configurations {$(n_4)$} exist for all {$n\geq17$}.
\newblock {\em European J.~Combin.}, 30(8):1778--1785, 2009.

\bibitem[BP13]{BokowskiPilaud-topologicalConfigurations}
J\"urgen Bokowski and Vincent Pilaud.
\newblock Enumerating topological $(n_k)$-configurations.
\newblock To appear in \emph{Comput. Geom.}, 2013.

\bibitem[BS05]{BokowskiSchewe1}
J{\"u}rgen Bokowski and Lars Schewe.
\newblock There are no realizable {$15_4$}- and {$16_4$}-configurations.
\newblock {\em Rev.~Roumaine Math.~Pures Appl.}, 50(5-6):483--493, 2005.

\bibitem[BS13]{BokowskiSchewe2}
J{\"u}rgen Bokowski and Lars Schewe.
\newblock On the finite set of missing geometric configurations {$(n_4)$}.
\newblock {\em Comput. Geom.}, 46(5):532--540, 2013.

\bibitem[Gr{\"u}00]{Grunbaum2}
Branko Gr{\"u}nbaum.
\newblock Connected {$(n_4)$} configurations exist for almost all {$n$}.
\newblock {\em Geombinatorics}, 10(1):24--29, 2000.

\bibitem[Gr{\"u}02]{Grunbaum3}
Branko Gr{\"u}nbaum.
\newblock Connected {$(n_4)$} configurations exist for almost all {$n$}---an
  update.
\newblock {\em Geombinatorics}, 12(1):15--23, 2002.

\bibitem[Gr{\"u}06]{Grunbaum4}
Branko Gr{\"u}nbaum.
\newblock Connected {$(n_4)$} configurations exist for almost all
  {$n$}---second update.
\newblock {\em Geombinatorics}, 16(2):254--261, 2006.

\bibitem[Gr{\"u}09]{Grunbaum1}
Branko Gr{\"u}nbaum.
\newblock {\em Configurations of points and lines}, volume 103 of {\em Graduate
  Studies in Mathematics}.
\newblock American Mathematical Society, Providence, RI, 2009.

\bibitem[OC12]{PaezOsunaSanAugustinChi}
Octavio~P{\'a}ez Osuna and Rodolfo~San~Agust{\'{\i}}n Chi.
\newblock The combinatorial {$(19_4)$} configurations.
\newblock {\em Ars Math. Contemp.}, 5(2):231--237, 2012.

\end{thebibliography}

\end{document}